# Radiation Peculiarities in Chiral Photonic Crystals

A.H. Gevorgyan [1], K.B. Oganesyan [2,3*], E.A. Ayryan [3], Michal Hnatic [4,5,6], Yuri V. Rostovtsev [7]

[1] Yerevan State University, Yerevan, Armenia
[2] Alikhanian National Lab, Yerevan Physics Institute, Yerevan, Armenia
[3] LIT, Joint Institute for Nuclear Research, Dubna, Russia
[4] Faculty of Sciences, P. J. Safarik University, Park Angelinum 9, 041 54 Kosice, Slovakia
[5] Institute of Experimental Physics SAS, Watsonova 47, 040 01 Kosice, Slovakia ˇ
[6] BLTP, Joint Institute for Nuclear Research, Dubna, Russia
[7] University of North Texas, Denton, TX, USA

[*] bsk@yerphi.am

The peculiarities of defect modes and reflection spectra of chiral photonic crystals (CPC) with an isotropic defect for various thicknesses of the defect layer was investigated. The peculiarities of the spectra of photonic density of the state (DOS) and those of the intensity at the center of the defect were investigated, too. It was shown that there is one more possibility of tuning of laser emission – by the change of the defect layer thickness – in CPCs doped with laser dyes (with resonance atoms) and with an isotropic defect. It was shown that the given system can work as a tunable narrow-band filter (or mirror) with tunable wavelength width and tunable location of frequency of the complete transmittance (reflection) band. The radiation peculiarities of these systems were discussed. It was shown that in certain conditions an anomalous strong radiation can be observed for defect modes, and low-threshold laser generation is possible, again, for these modes. It was also shown that radiation suppression is possible for defect modes. The peculiarities of group velocity, group velocity dispersion and nonreciprocity were studied.

## I. Introduction

The cholesteric liquid crystals (CLC) are the most well-known representatives of one dimensional (1D) chiral photonic crystals (CPC), because they can spontaneously self organize their periodic structure, and their photonic band gap (PBG) can be tuned in wide frequency intervals. They also possess other wonderful optical properties. Recently these media have drawn great interest to them due to their possibility of low-threshold laser generation at edges of their PBG (predicted by Dowling et al [1] and experimentally established by Kopp et al [2]). Vigorous investigations in this area have been going on up to now (for instance, see the references cited in [3]). The main feature of a CPC is its possession of the circular Bragg phenomenon, meanwhile a circularly polarized normally incident plane wave with specific handedness is strongly reflected in a certain wavelength regime, whereas a similar plane wave but of reverse handedness is not. This polarization-discriminatory filtering characteristic of a CPC is very attractive in optical technology.

Recently great interest is drown to the artificial CPCs, too [4-10]. Besides, recently the CPC having various types of defects have been considered from the point of view of generating additional resonance modes in them and of investigating of possibilities of low-threshold laser generation for these modes (again, see the references cited in [3]). It is to be noted that the CLC with a defect in the structure possess a number of peculiarities which the isotropic 1D PCs lack (see below). In papers [3, 11-13] the CPCs with an anisotropic defect were considered, both for the cases of normal light incidence [3, 11, 13] and oblique light incidence [12]. In [13] the peculiarities of the polarization characteristics, of photonic density spectra of states and light intensity at the defect centre, as well as *Q*-factor of the subject system were investigated like [41-108].

In this paper we theoretically investigate some new peculiarities of the defect modes in CPC with an isotropic defect and have found out different features of such a system.

## II. Results and Discussion

Let us consider light propagation from the left through a multi-layer system, CPC(1) – Isotropic Dielectric Layer (IDL) – CPC(2). This problem was first considered in [14]. Here we consider some new interesting properties of this system, which can have important applications. The method of solution of this problem is described in detail in [3,11-13], therefore we immediately pass on to the obtained results.

We discuss and compare in detail the following four cases:

1. Absence of defects.
2. A super thin defect ($d^d << \lambda$) – light accumulation in the defect takes place in this case, and also low-threshold laser generation and amplified non-linearity are possible.
3. A comparatively thick defect.
4. A thick defect ($d^d >> \lambda$) – the number of defect modes is increased in this case.

In Figure 1 the spectra of reflection coefficient (the first column), photonic density of states (the second column) and light intensity at the defect centre (the third column) when there is no defect (the first row) and for various defect layer thicknesses (the second, third and fourth rows) are presented.

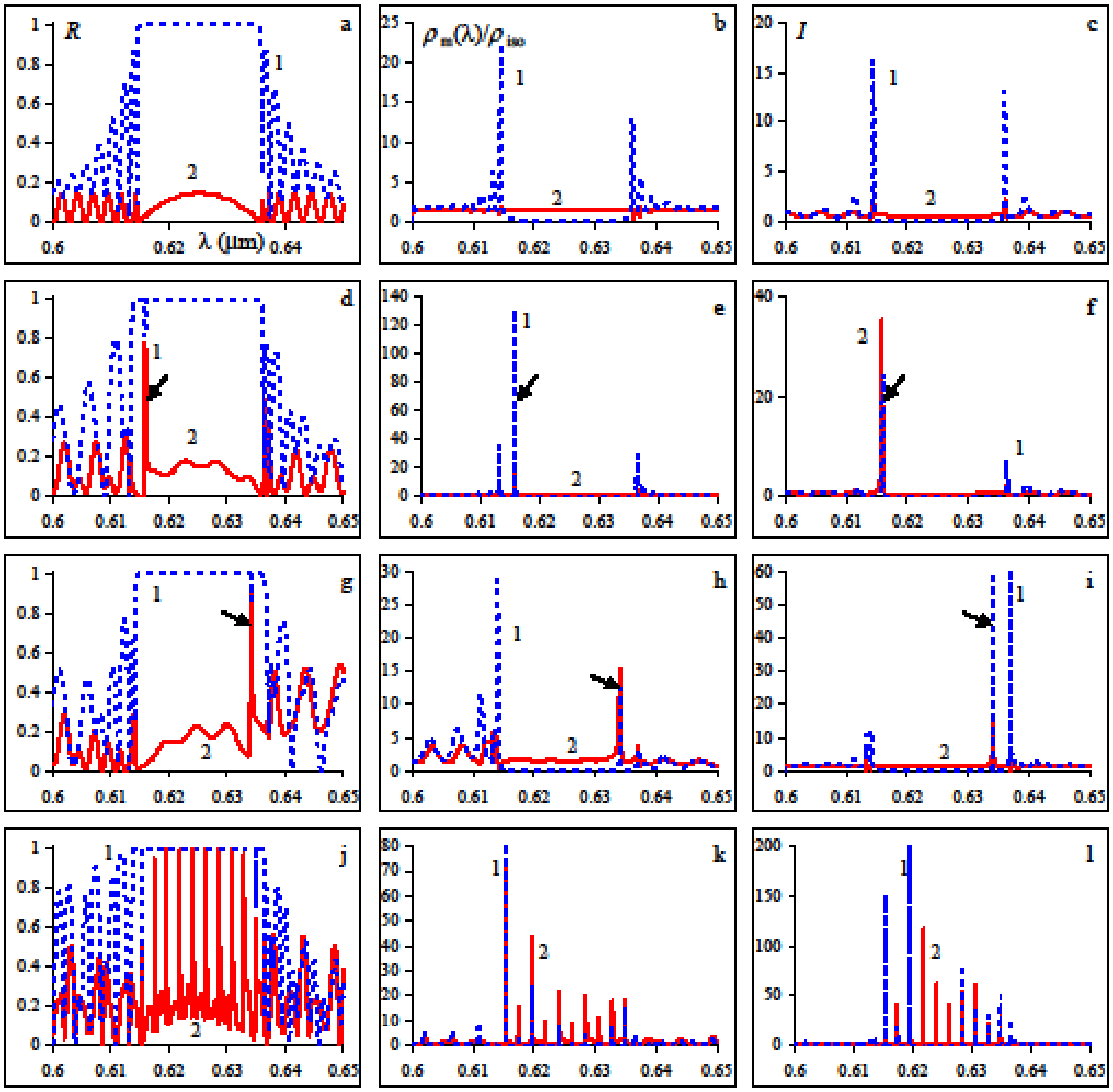

Fig. 1. The reflection coefficient spectra (the first column), the spectra of the relative DOS, $\rho_m(\lambda)/\rho_{iso}$, (the second column), the spectra of the light intensity at the defect center, $I=\left|E\left(z=(d+d^d)/2\right)\right|^2$ (the third column) for various defect layer thicknesses. $d^d$=0 (in the first row), $d^d$=0.016μm (in the second row), $d^d$=1.3μm (in the third row) and $d^d$=50μm (in the fourth row). The light incident on the system has polarization coinciding with the first EP (curve 1) and the second EP (curve 2). The CPC helix is a right one. The CPC parameters are as follows: the main values of dielectric permittivity tensor are $\varepsilon_1$=2.29 and $\varepsilon_2$=2.143; the helix pitch is σ=0.42μm; the CPC thickness is $d$=100σ, and the defect dielectric permittivity is $\varepsilon^d$ = 2.7. The defect layer is at the center of the CPC.

The incident light has polarization coinciding with the first eigen polarization (EP) diffracting on the CPC structure) (curve 1), and with the second eigen polarization (curve 2). Eigen polarizations are the two polarizations of the incident light, which do not change when light transmits through the system. These polarizations mainly coincide with those of eigen waves arisen in the system (for more details on EP in CPC see in [13]). The CPC helix is a right one.

Comparison of the reflection spectra in Fig. 1a and d shows that the presence of a thin defect in CPC structure leads to arising a defect mode in PBG (shown by an arrow in Fig. 1). It manifests itself in the form of a dip in the reflection spectrum of the light with the first EP (diffracting EP), and in the form of a peak in the reflection spectrum of the light with the second EP (non diffracting EP) (Fig. 1d). As it is seen from Fig. 1d, the reflection is the same for both EP at this defect mode and, as the calculations show, the polarization plane rotation and polarization ellipticity vanish at this defect mode. Thus, though the system is an anisotropic and inhomogeneous one, single refringence takes place at the defect mode (just as for the isotropic medium layer case).

The defect mode has either a donor or acceptor character depending on the defect layer thickness. The defect wavelength increases from the minimum to the maximum of PBG with the increase of the defect's optical thickness. Besides, first two defect modes appear nearby the band gap, and then the longer wave mode vanishes as the defect layer thickness increases, and the short wave mode shifts to the long wave region [14].

In the case of an anisotropic defect, an additional phase difference appears which leads to significant peculiarities if the defect layer thickness increases. In particular, the defect mode half-width becomes dependent on the defect layer thickness and, for instance, if $d^d \sim \lambda / 2\left(n_e^d - n_o^d\right)$, i.e. if the defect is a half wave plate, complete reflection (non selective in respect to polarization) takes place in PBG [3, 11]. In the case of an isotropic defect, the change of the half-width of the defect mode is insignificant if the defect thickness is changed. This is an important advantage in certain conditions.

Indeed, in the case of an anisotropic defect, one can obtain abundant light accumulation, a large *Q*-factor for the defect mode, and low-threshold laser generation – for a very thin planar defect layer – which arouse practically almost unsurpassable technical difficulties for designing optical devices . In the case of an isotropic defect, one can also obtain abundant light accumulation, a large *Q*-factor for the defect mode, and low-threshold laser generation for comparatively greater thicknesses of the defect layer (see, for instance, Fig. 1g). In the case of very large defect layer thickness, $d^d >> \lambda$, the number of defect modes increases and again, as our calculations show, single refringence takes place at these modes. The frequency location and number of these modes can be varied, changing the defect layer thickness.

As it was said in Introduction, the CPC doped with laser dies (with resonance atoms) can be used to design lasers with feedback, and in certain cases to design non-mirror ones. In the amplifying media (in particular, in CLCs doped with fluorescent guest molecules with the peak of fluorescent radiation either located in PBG, or involving the PBG) the PBG essentially influences the radiation spectrum. Inside the PBG the wave is damping and it evanescently leads to suppression of the spontaneous radiation. The reason of this is that the DOS vanishes in this case and, as it is known [15, 16], that the spontaneous radiation intensity is proportional to DOS, so that the spontaneous radiation intensity also vanishes. Indeed, according to the theory developed in [15, 16], the spontaneous radiation intensity inside the layer and at the point $z$ is defined by the expression:

$$p(\lambda, z) = \frac{\rho_m(\omega)}{\rho_{iso}} \frac{\left\langle |\mathbf{d}|^2 \right\rangle |E_m(z)|^2}{U(k)}, \tag{1}$$

where $\rho_m$ and $E_m(z)$ are the DOS and electric field of the $m^{\text{th}}$ EP, $\rho_{\text{iso}}$ is the DOS of the homogeneous isotropic layer with the refraction coefficient, $n=\bar{n}$, $\left\langle |\mathbf{d}|^2 \right\rangle$, averaged in respect to the orientation distribution of dipole momentum transitions, $U(k)$ is the total electric energy accumulated in the CPC. Let's note, that the most general formula for the spontaneous emission rate within a photonic crystal is provided by Vats and coauthors [17].

At the PBG borders a sharp increase of the spontaneous radiation life time, $\tau_{s,}$ takes place ($\tau_s$ decreases with oscillations outside the PBG) leading to strong increase of the forced radiation. In CPC, the total electric field of the diffraction mode is linearly polarized – perpendicular (parallel) to the director – at the short wave (long wave) border of the PBG. Therefore, if dipole momentum transitions are highly orientated along the director, the radiation intensity increases at the short wave border, providing laser generation. In these conditions, the laser generation threshold energy essentially decreases, and the radiation increases.

Below we investigate some spectra peculiarities of DOS, and the light intensity spectra at the defect centre. We investigate tuning possibilities of the laser radiation wavelength by changing the defect layer thickness.

DOS is reverse to the group velocity and is defined by the expression [18, 19]:

$$\rho(\omega) \equiv \frac{dk}{d\omega} = \frac{1}{L}\frac{\frac{du}{d\omega}v - \frac{dv}{d\omega}u}{u^2+v^2}, \tag{2}$$

where $d$ is the CPC thickness, and $u$ and $v$ are the real and imaginary parts of the transmittance coefficient.

In Fig. 1 the spectra of $\frac{\rho_m(\lambda)}{\rho_{iso}}$ (the second column) and $I = \left|E\left(z = (d + d^d)/2\right)\right|^2$ (the third column) are presented for the same defect layer thickness. As it can be seen from the Figure and as it could have been anticipated, being rather well-known (see, for instance, [15-17]), the DOS has peaks at the PBG borders if there is no defect in the structure. In the presence of a defect, large peaks of DOS are observed at the defect mode (Fig. 1e, h), which are accompanied with a partial suppression of DOS at the PBG borders.

Let us note that investigations of light energy distribution in the system is important from another point of view, namely, from the viewpoint of light energy accumulation possibility.

Thus, the presented results demonstrate the existence of one more possibility of tuning the laser generation wavelength in CPC, namely, changing the defect layer thickness.

Now we pass on to the investigation of radiation peculiarities of CPC with an isotropic defect.

Let the CPC and the defect layer be doped with dye molecules. If the wavelength is a pumping one, this system is amplifying, i.e. the system is a planar resonator with an active element. The presence of dye molecules leads to changes of local refraction coefficients of the system. In this case the effective imaginary parts of the effective local refraction coefficients both of CLC ($n''_{1,2}$) and of the isotropic defect ($n''^d$) are negative ($n_{1,2} = n'_{1,2} + in''_{1,2}$ and $n^d = n'^d + in''^d$). If there is absorption (in this case the above-said coefficients of the imaginary parts are positive), the quantity $A$=1–($R$+$T$) characterizes the absorbed by the system light energy ($R$ and $T$ are the reflection and transmission coefficients respectively) and $A$<1, and in an amplifying medium the radiation of the system is characterized by $|A|$.

Now we characterize the degree of ordering of dipole momentum transitions of guest molecules through the order parameter $S_d$, which is defined through the averaged $\cos\vartheta$:

$$S_d = \frac{3}{2}\langle\cos\vartheta\rangle - \frac{1}{2}, \tag{3}$$

where $\vartheta$ is the angle between the CPC optical axis and the dipole momentum transitions of guest molecules. The possible maximum value of the order parameter, $S_d$=1, corresponds to the ideal case of dipole momentum transitions orientation along the local optical axis. The case $S_d$=0 corresponds to the isotropic distribution, and the case $S_d$=–0.5 corresponds to the isotropic distribution of dipole momentum transitions in the plane perpendicular to the local optical axis. In the linear optics approximation, the obtained expressions describe both the amplification and generation regimes.

Let us now assume that $n''_{1,2}$ and $n''^{d}$ are negative, i.e. the system is an amplifying one. If the quantities $|n''_{1,2}|$ and $|n''^{d}|$ are sufficiently small, then the waves coming out of the system will exist only when there is an incident on the system wave, and in this case the quantity $|A|$ characterizes an amplifying system. But if the imaginary parts of $|n''_{1,2}|$ and $|n''^{d}|$ reach certain values, $R$ and $T$ undergo sharp changes at a certain wavelength (depending on the parameters of the system) and tend to infinity, and the amplitudes of the waves coming out of the system can be different from zero even for the zero amplitude of the incident wave. In such situation the amplitudes of both reflected and transmitted waves cannot be defined from the linear problem, and a non-linear problem must be solved. And if this is done, then requiring non-triviality for the found amplitudes of the reflected and transmitted waves for the zero incident wave, one can define the so called laser mode frequencies and the respective values of amplifications, i.e. the minimum threshold amplification at which laser generation takes place (see the analogous considerations for a periodic layered structure and for a chiral periodic medium in [20-23]). As it was shown in [22, 23], the problem with the above-said condition has an analytical solution for the CLC with an ideal periodic structure and for small values for the parameter $d\mathrm{Im}k$ ($d$ is the CLC thickness, $k$ is the wave number of the diffracting mode in the rotating frame). In the general case (and in our case, too), it is difficult to obtain an analytical expression for the laser mode frequencies and the respective values of amplifications. Nevertheless, it is possible to get laser mode frequencies and the respective threshold values of amplifications from the sharp maximums of the quantity $|A|$.

In Fig. 2 the spectra of $\ln(|A|)$ for $S_d$=0 (the first column) and for $S_d$=1 (the second column) and for $S_d$= -0.5 (the third column) are presented. The first row corresponds to the case of the super thin defect ($d^d$=0.016μm), the second row corresponds to the case of a comparatively thicker defect ($d^d$=1.3μm), and the third one presents the case of the thick defect ($d^d$=50μm). The light incident on the layer has polarization coinciding with the first (diffracting) EP (curve 1), and with the second EP (curve 2).

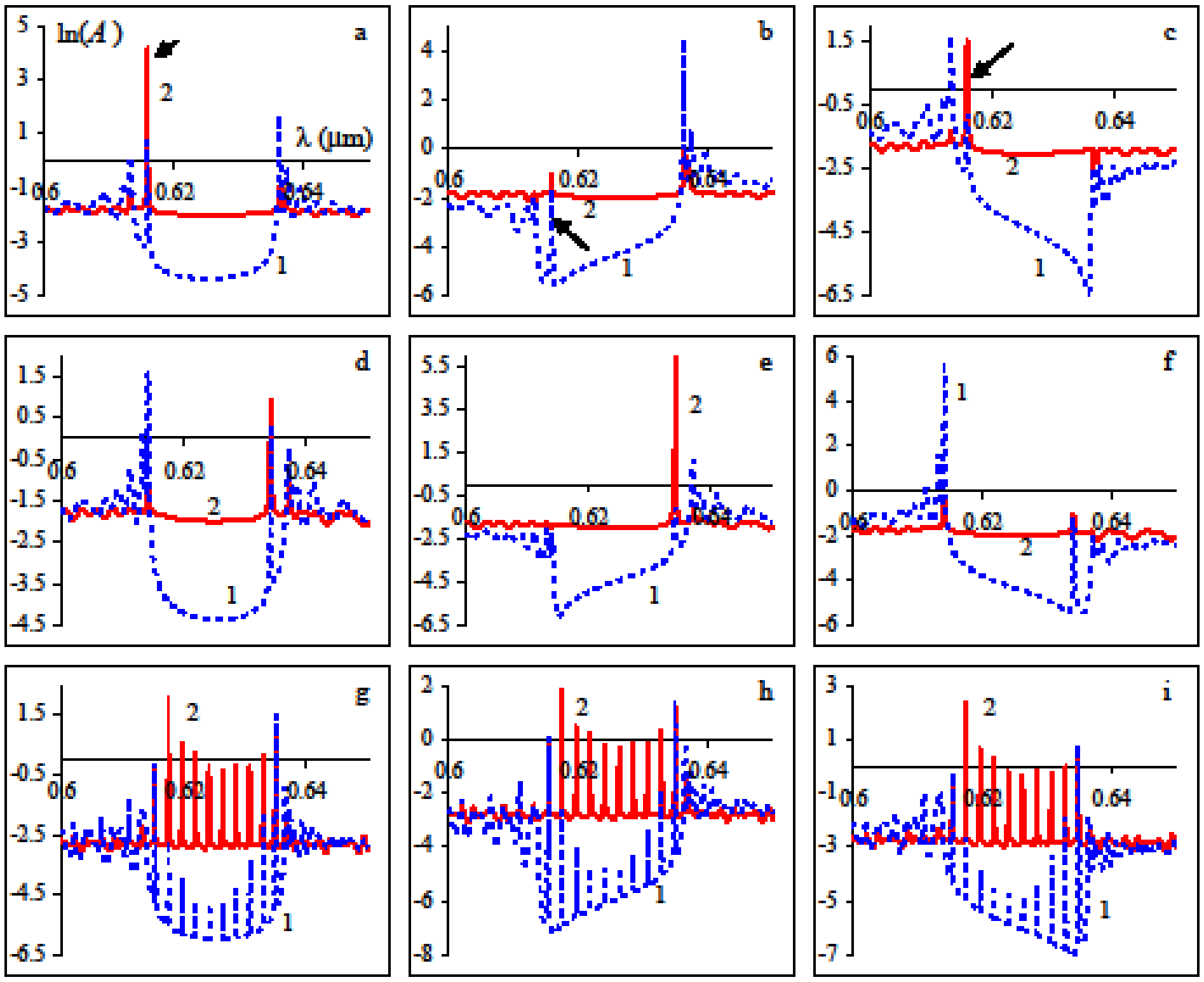

Fig. 2. The spectra of $\ln(|A|)$ at $S_d$=0 (the first column), at $S_d$=1 (the second column) and at $S_d$= -0.5 (the third column). The first row respects to the case of the super thin defect ($d^d$=0.016μm: $\mathrm{Im}\varepsilon_m = \frac{\mathrm{Im}\varepsilon_1 + \mathrm{Im}\varepsilon_2}{2} = -0.0005, \mathrm{Im}\varepsilon^d = -0.0005$); the second row respects to a comparatively thicker defect ($d^d$=1.3μm: $\mathrm{Im}\varepsilon_m = \frac{\mathrm{Im}\varepsilon_1 + \mathrm{Im}\varepsilon_2}{2} = -0.0005, \mathrm{Im}\varepsilon^d = -0.0005$); and the third one is the case of a thick defect ($d^d$=50μm: $\mathrm{Im}\varepsilon_m = \frac{\mathrm{Im}\varepsilon_1 + \mathrm{Im}\varepsilon_2}{2} = -0.0001, \mathrm{Im}\varepsilon^d = -0.0001$). the incident light has a polarization coinciding with the first EP (curve 1) and with the second EP (curve 2). The CPC helix is a right one. The CPC parameters are the same as in Fig. 1 and $\mathrm{Im}\varepsilon_m = \frac{\mathrm{Im}\varepsilon_1 + \mathrm{Im}\varepsilon_2}{2} = -0.0005, \mathrm{Im}\varepsilon^d = -0.0005$.

As it is seen from Fig. 2a, in PBG (covering the range λ=0.6148μm to λ=0.6356μm for the given parameters of the problem), a resonance decrease of radiation (of $\ln(|A|)$) is observed for the light with diffracting EP, which is analogous to diffraction suppression of absorption [24-26]. But at the PBG borders an anomalous strong radiation takes place, which is analogous to the anomalous absorption nearby the PBG borders [24-26]. And at the defect mode with the central wavelength, λ=0.6159μm, i.e. nearby the shortwave borders of the PBG (the defect modes in the Figures are shown by arrows) $\ln(|A|)$ has a sharp peak, and an anomalous strong radiation takes place here. This means that low threshold laser generation is possible for this mode. Furthermore, the comparison of these results with the analogous results for the case of the absence of the defect shows that the presence of the defect leads to certain suppression of radiation at the PBG borders. For $S_d=1$ ($S_d = -0.5$) practically complete suppression of radiation takes place at the shortwave (longwave) border of PBG and, reversely, an anomalous strong radiation is observed at the longwave (shortwave) borders of PBG.

These effects are the analogues to the effects of anomalous strong and anomalous weak light absorption observed in CPCs for anisotropic absorption [24-26]. Due to the fact that for the given parameters of the problem, the defect mode is nearby the shortwave border of the PBG, a partial suppression of radiation is observed at the defect mode for $S_d=1$ (Fig, 2b).

In the case of a comparatively thick defect ($d^d$=1.3μm, with the central wavelength of the defect mode nearby the longwave border and at λ=0.63407μm) the strongest radiation at the defect mode is observed for $S_d$=1. For $S_d$=–0.5, the suppression of radiation is observed both at the longwave border of PBG and at the defect mode.

In the case of the thick defect, the number of those defect modes is increased, which have wavelengths defined approximately through the well-known expression for Fabri-Perrot resonator:

$$\lambda_{m+1} = \frac{\lambda_m}{1 - \dfrac{\lambda_m}{2d^d n^d}}, \qquad (4)$$

where $n^d$ is the refraction coefficient of the defect. As it is seen from Fig. 2, an anomalous radiation takes place at the defect modes and a multi-mode generation is possible also for the thick defect.

In the Fig. 3, the same dependences as in Fig. 2 are presented, but for greater (10 times) values of Im($\varepsilon_m$) and Im($\varepsilon^d$) ($\varepsilon_m = \frac{\varepsilon_1 + \varepsilon_2}{2}$, $\varepsilon_1$, $\varepsilon_2$ are the principal values of the CPC dielectric permittivity tensor, $\varepsilon^d$ is the dielectric permittivity of the defect layer). Comparison of Fig.2a with Fig.3a shows that if Im($\varepsilon_m$) and Im($\varepsilon^d$) increase, the radiation mode is suppressed at the defect and, reversely, the radiation nearby the PBG borders is increased. To understand this peculiarity, one has to take into account the following. As it is known (see [25]), in the case of the CLC with an ideal periodic structure, the frequency minimums of the reflection coefficient are defined by the condition:

$$\frac{\omega}{c}\sqrt{\varepsilon_m}\sqrt{1+\frac{2}{4}\left(\frac{4\pi c}{\omega\sigma\sqrt{\varepsilon_m}}\right)^2-\sqrt{\left(\frac{4\pi c}{\omega\sigma\sqrt{\varepsilon_m}}\right)^2+\delta^2}}=\pi m, \tag{5}$$

where $\gamma=\frac{\operatorname{Im}\varepsilon_m}{\operatorname{Re}\varepsilon_m}$, $\delta=\frac{\varepsilon_1-\varepsilon_2}{\varepsilon_1+\varepsilon_2}$, σ is the CPC helix pitch, $m$=1,2,3….

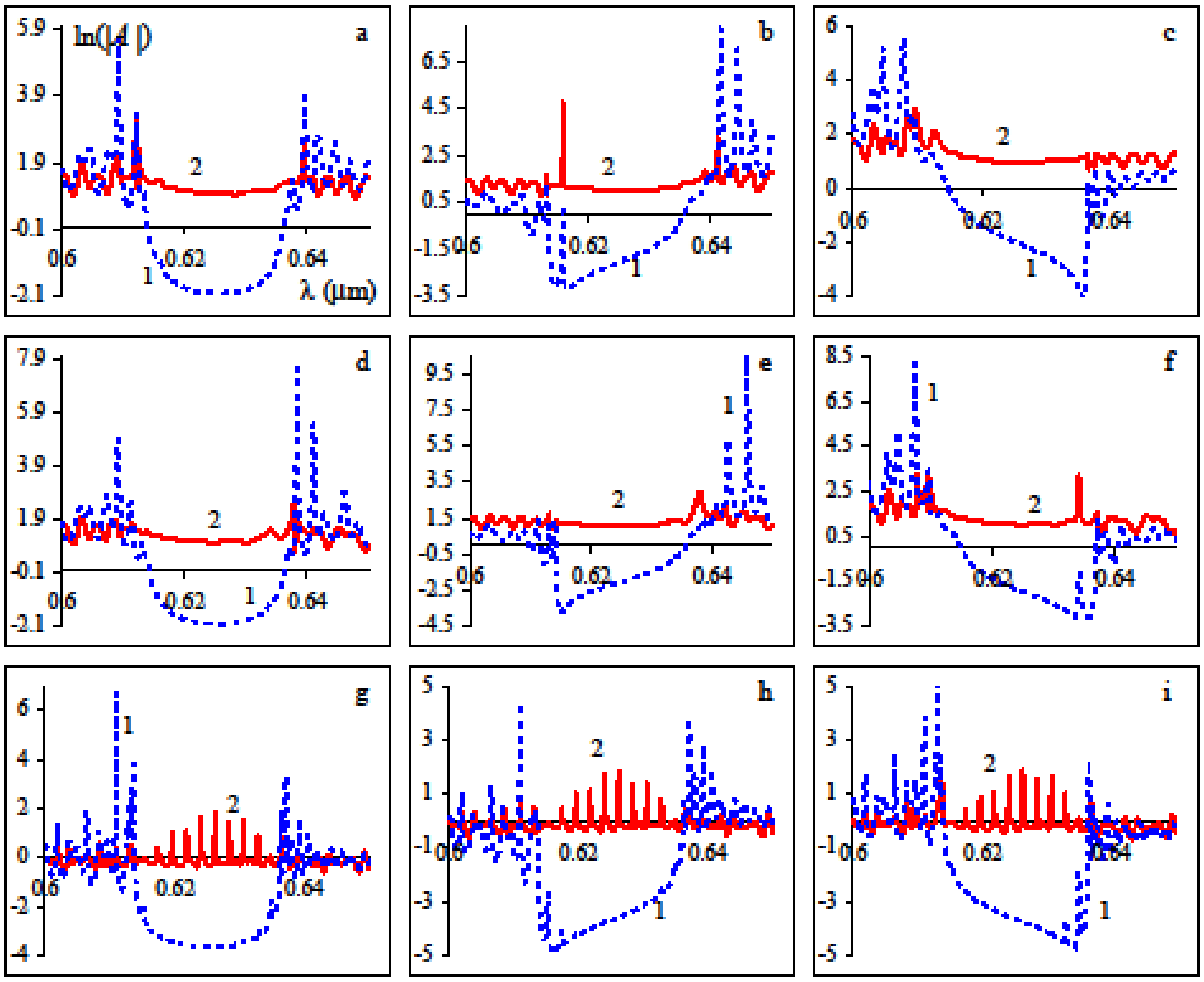

Fig. 3. The same as in fig. 2, and the dependences are for tenfold values of Im($\varepsilon_m$) and Im($\varepsilon^d$).

For weak absorption (radiation) and small $d$Im$k$ the reflection and transmission coefficients are defined by the following expressions for the said frequencies:

$$R=\frac{\left(a^3\gamma\right)^2}{\left(\left(\pi m\right)^2+a^3\gamma\right)^2},\quad T=\frac{\left(\pi m\right)^4}{\left(\left(\pi m\right)^2+a^3\gamma\right)^2},\tag{6}$$

where $a=\dfrac{\pi\delta d}{\sigma}$, $\gamma=\dfrac{\operatorname{Im}\varepsilon_m}{\operatorname{Re}\varepsilon_m}$, $\delta=\dfrac{\varepsilon_1-\varepsilon_2}{\varepsilon_1+\varepsilon_2}$, σ is the CPC helix pitch, *m*=1,2,3…. In the case of amplification, γ is negative and, consequently, if:

$$\gamma=-\frac{\left(m\pi\right)^2}{a^3}=-\frac{\left(m\pi\right)^2}{\left(\pi\delta d/\sigma\right)^3},\tag{7}$$

then the reflection and transmission coefficients divergence. As it is shown in [23, 24], for small values of $|\gamma|$, the laser generation frequencies are defined by condition (5) (as is shown in [23, 24], the solutions of the equation which is the condition of existence of non-trivial amplitudes of solutions for reflected and transmitted waves –for the case if the incident wave has zero amplitude – coincide with the solutions of (6)), and the corresponding threshold values of γ are defined from condition (7).

As it is shown in [23, 24], and as it is seen from (7), the minimum threshold value of $|\gamma|$ corresponds to the mode with *m*=1, i.e. to the mode nearest to the PBG laser mode, and the threshold value of γ increases if *m* is increased. If the CPC structure has a defect, defect modes are added to the laser modes defined by (5), and one must ascribe the least order number exactly to the defect modes. Thus, the least threshold value of $|\gamma|$ corresponds exactly to the defect mode for the case of the thin defect (see, for instance, Fig.2a), or to the defect modes for the case of the thick defect. The modes with large *m* correspond to large threshold values of γ (compare Fig.3a with Fig.2a).

Let us note once more that the subject problem on transmittance of radiation through a planar resonator with an active element and a constant amplification coefficient is not adequate to a real process. The amplification coefficient decreases as the intensity of the propagating through the medium wave increases. This is connected with the peculiarities of appearing of the inverse state – at very large energies accumulated in the laser active element, the speed of forced transitions prevails over the pumping speed. This leads to a sharp decrease of the difference of population of exited and ground states, which, in its turn, leads to a decrease of the amplification coefficient and, consequently, to the saturation of intensity [27]. And, as the interaction of radiation with the amplifying medium becomes neither linear, nor stationary, the linear approximation cannot be applied to the subject problem. Nevertheless, the presented results give much information on radiation peculiarities and laser generation in the CLC with a defect and, in particular, on defect laser modes and threshold values of amplification coefficient.

### 3. The Group Velocity. Group Velocity Dispersion

As it is known, the light group velocity can decrease or increase in dispersive media nearby resonances [28,29]. The sharp decrease of the light group velocity can even lead to a complete stopping of light signals, which has been recently demonstrated in experiments with various atomic and solid systems.

Again, recently investigations of the light group velocity change in 1D Photonic Crystals (PC) are of great interest. The super-high and super-low light group velocities have been also demonstrated in experiments [30-35], as well as in many other works (see, for instance, [36] and its references). In particular, these investigations are very important for the optical communication techniques. Potential applications of slow-light and superluminal light include controllable optical delay lines, optical buffers, true time delay for synthetic aperture radars, and cryptography and imaging in the quantum information field. Superluminal light propagation is interesting for scientific community specifically in the context of information velocity, since superluminal signal velocities can be achieved without violating the Einstein causality [37].

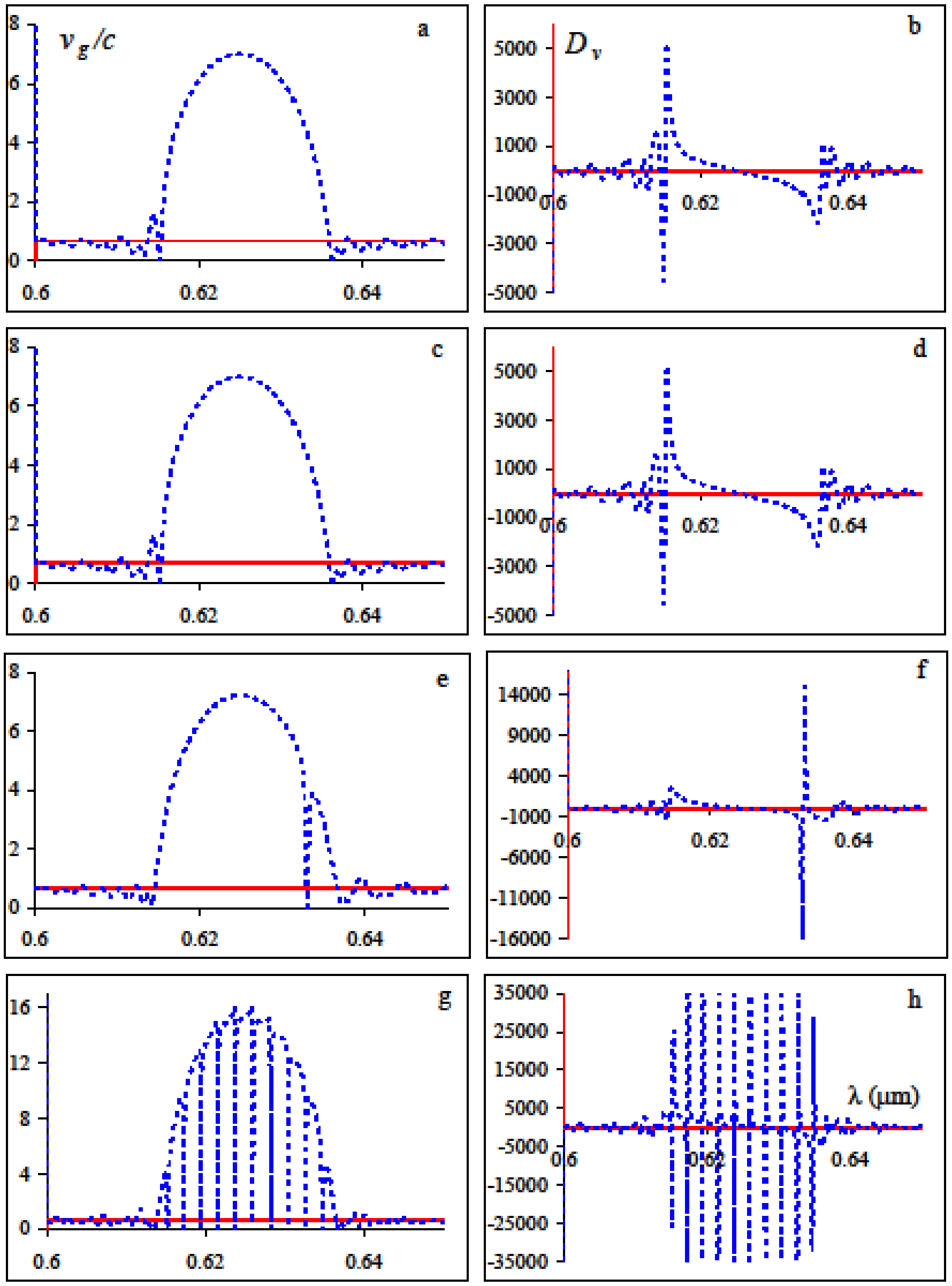

Fig.4. The spectra of the relative light group velocity, $v_g/c$, (the first column) and of the group velocity dispersion, $D_v$, (the second column) for various defect layer thicknesses. $d^d$=0 (in the first row), $d^d$=0.016μm (in the second row), $d^d$=1.3μm (in the third row) and $d^d$=50μm (in the fourth row). The CPC parameters and the notations are the same as in Fig. 1.

We are investigating the influence of a defect in chiral PC on the group velocity and its dispersion $\left( D_v = \frac{1}{c}\frac{dv_g}{d\lambda} \right)$. The importance of this investigation, as it is known, is due to the fact that dispersion is the key factor for preserving the signal form, and this can provide the lowest level of the code error in communication cables.

In Fig.4 the spectra of the relative light group velocity, $v_g/c$, (the first column) and of the group velocity dispersion, $D_v$, (the second column) are presented for the two orthogonal circular polarizations, when there is no defect (the first row) and for various defect layer thicknesses (the second, third and fourth rows).

As it is seen from this Figure, a super propagation of light is observed in PBG, meanwhile $v_g/c \to 0$ at the PBG borders, i.e. a large increase of the time delaying of light propagation through the system takes place. The group velocity dispersion, $D_v$, oscillates outside the PBG, and it decreases monotonously in the PBG. Also, a strong resonance change both of the group velocity and its dispersion, i.e. effects of fast-light and slow-light are observed at defects modes. Such behavior of the above said factors can be very important, because significant transmittance takes place at these modes.

It is known that if the momentary frequency decreases from the beginning of the light signal to its end, then in the medium where the group velocity linearly decreases depending on the wavelength, the head and tail parts of the signal start to come near to each other, and a compression of the frequency-modulated signal in a dispersing medium is observed. Otherwise (in the reverse case), the wave packet is spread. As it is seen from Fig. 4, the PC with a changing spatial modulation period can be used for compressing or spreading of light signals.

### 4. The Optical Nonreciprocity

The loss of the reciprocity in the optical system can lead to an important new class of optical devices, such as optical isolators, which are critical for the development of photonic systems. Traditionally, creation of non- reciprocity devices is based on magneto-optical effects, but recently new non-reciprocity mechanisms have been found [38-41].

An all-optical diode is a nonreciprocal device that – in ideal case and for specific wavelength or wavelength band – allows total light transmission in the forward direction ($T^+=1$), and totally inhibits light propagation in the backward direction ($T^-=1$), yielding a unitary contrast C= $(T^+ - T^-)/(T^+ + T^-)=1$. All-optical diodes are widely considered to be the key components for the next generation of all-optical signal processing, in complete analogy with electronic diodes which are widely used in computers and etc. for the processing of electric signals. Replacing relatively slow electrons with photons as carriers of information would substantially increase the speed and the bandwidth of telecommunication systems, which will lead to a real revolution of the telecom industry.

Due to the crucial technological implications, such unidirectional propagation (‘diode action’) has been studied experimentally by several groups, using the most diverse schemes and experimental techniques [38, 43, 44].

It is to be expected that the use of a CPC with an asymmetric defect location will break the forward/backward symmetry, which we demonstrate below.

In Fig. 5a the reflectance coefficient dependence on the wavelength for the asymmetric defect location is presented, and the dependence of C on the wavelength is presented in Fig. 5b.

As it is seen from the figure, C tends to 1 for certain wavelengths (for instance, at $\lambda=0.611\mu m$, C=0.91), and this shows that this system can be used as an all-optical diode.

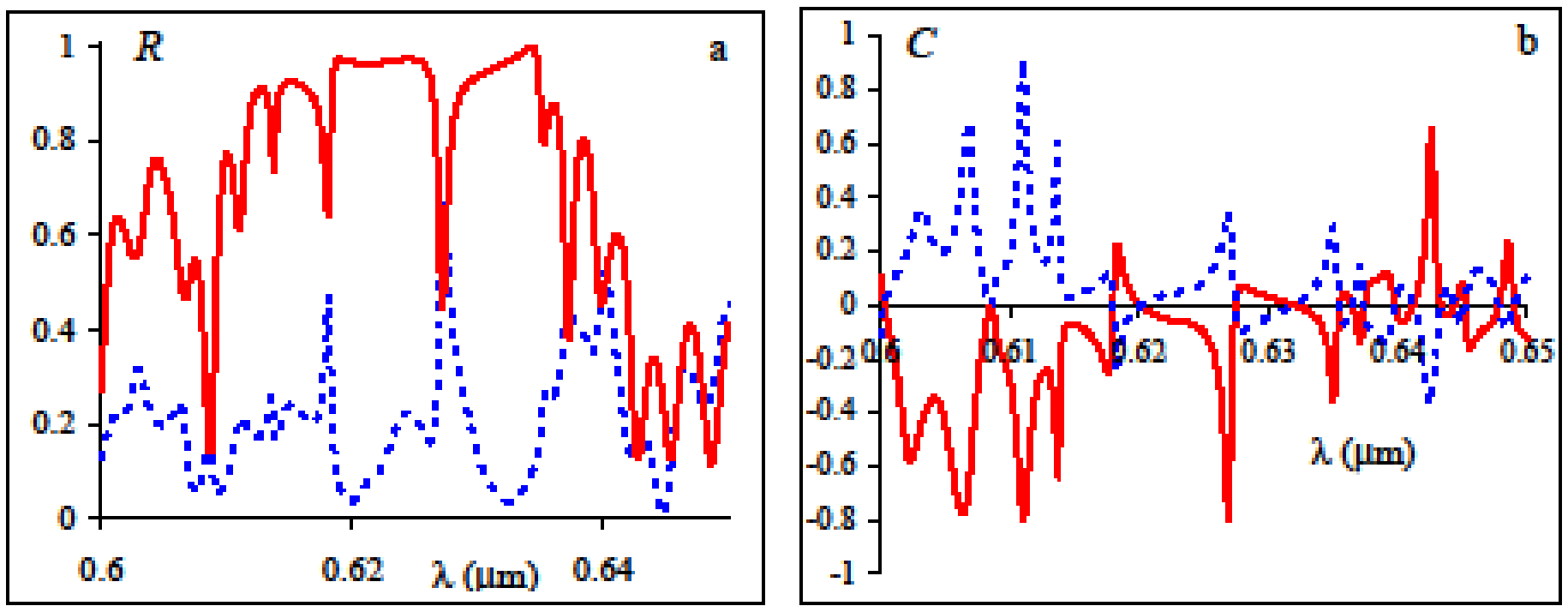

Fig.5. The reflection coefficient spectra (a), the spectra of the *C* (b). The ratio of the layer thicknesses of CPC on defects left and right is 1/6. The light incident on the system has: (a) right circular polarization (red solid line) and left circular polarization (blue dashed line): (b) linear polarization along *x* axis (red solid line) and along *y* axis (blue dashed line). The CPC helix is a right one. The defect layer thickness is: $d^d$=10μm. The other parameters are the same as in Fig. 1